# Optical Screening of Citrus Leaf Diseases Using Label-Free Spectroscopic Tools: A Review

Saurav Bharadwaj[1,2], Akshita Midha[2], Shikha Sharma[3], Gurupkar Singh Sidhu[4] and Rajesh Kumar[2*]

[1]Agriculture & Water Technology Development Hub, Indian Institute of Technology Ropar, Punjab-140001, India.
[2]Department of Biomedical Engineering, Indian Institute of Technology Ropar, Rupnagar, Punjab-140001, India.
[3]Department of Plant Pathology, College of Agriculture, CSK Himachal Pradesh Agricultural University, Palampur, Himachal Pradesh-176062, India.
[4]School of Agricultural Biotechnology, College of Agriculture, Punjab Agriculture University, Ludhiana, Punjab 141004, India.

*email: rajeshkumar@iitrpr.ac.in

**Abstract**

Citrus diseases pose threats to citrus farming and result in economic losses worldwide. Nucleic acid and serology-based methods of detection such as polymerase chain reaction (PCR), loop-mediated isothermal amplification (LAMP), and immunochromatographic assays are commonly used but these laboratory tests are laborious, expensive and might be subjected to cross-reaction and contamination. Modern optical spectroscopic techniques offer a promising alternative as they are label-free, sensitive, rapid, non-destructive, and demonstrate the potential for incorporation into an autonomous system for disease detection in citrus orchards. Nevertheless, the majority of optical spectroscopic methods for citrus disease detection are still in the trial phases and, require additional efforts to be established as efficient and commercially viable methods. The review presents an overview of fundamental working principles, the state of the art, and explains the applications and limitations of the optical spectroscopy technique including the spectroscopic imaging approach (hyperspectral imaging) in the identification of diseases in citrus plants grown over a large area. The review highlights (1) the technical specifications of optical spectroscopic tools that can potentially be utilized in field measurements, (2) their applications in screening citrus diseases through leaf spectroscopy, and (3) discusses their benefits and limitations, including future insights into label-free identification of citrus diseases. Moreover, the role of artificial intelligence is reviewed as potential effective tools for spectral analysis, enabling more accurate detection of infected citrus leaves even before the appearance of visual symptoms by leveraging compositional, morphological, and chemometric characteristics of the plant leaves. The review aims to encourage stakeholders to enhance the development and commercialization of field-based, label-free optical tools for the rapid and early-stage screening of citrus diseases in plants.

## 1 Introduction

Citrus plants are evergreen trees and shrubs primarily grown in tropical and subtropical areas of Asia, Melanesia and Australia. Citrus fruits are rich in vitamins, minerals, dietary fibers, and phytochemicals, which are regularly consumed as part of a balanced diet and offer numerous health benefits to combat multiple diseases in human. However, citrus diseases, broadly classified as fungal, viral and bacterial diseases, significantly impact citrus crops, leading to substantial economic losses for citrus growers and consequently affecting the regional economies that depend on the citrus industry [1]. Alternaria brown spot in citrus fruits and leaves is a common disease caused by the airborne fungus *Alternaria alternata*. It develops foliar lesions on new tissues which eventually turns black spots into prominent yellow halos [2]. Anthracnose is another fungal disease that affects leaves and results in the colonization of senescent tissues. The diseases grow on dead wood in the canopy and are spread by rain. The sexual spores are deposited on the young leaves and immature fruits that lead to the germination of spores, forming a resting structure and remain dominant until an injury occurs [3]. The greasy spots are wind-dispersed sexual spores produced during wet periods from the decomposition of fallen leaves [4]. The yellow spots can be found on the adaxial surface of the leaves. The abaxial surface exhibits brown blisters that eventually turn black due to the deposition of sexual spores [5]. Citrus canker is caused by the Gram-negative bacterium *Xanthomonas axonopodis* when the bacteria enter the stomata on the leaves or other green parts of the plant (Fig. 1a). The infected stems and fruits are spotted with water-soaked brown margins and yellow rings around the lesions [6]. *Xylella fastidiosa* causes citrus variegated chlorosis (CVC) in citrus plants. It damages the vascular system of plants. It shows yellow veins on the adaxial surface of leaves and indicates the deficiency of zinc in infected leaves [7]. Huanglongbing (HLB), also known as citrus greening, is another devastating bacterial disease widely considered a significant threat to citrus cultivation worldwide. It is caused by the vector-transmitted pathogen *Candidatus Liberibacter* and can be identified through symptoms such as yellowing and mottling of leaves, premature leaf shedding, deterioration of twigs and decay of roots and rootlets [8] (Fig. 1b). Leaves remain small with yellow veins, deformed fruits, discoloured green areas and eventually death of the tree [9]. As a consequence of disease infestations, chlorosis - a chlorophyll deficit condition, may also simultaneously arise in citrus plants cause gradual diminishing of colour in leaf and results in contrasting the leaf veins. The affected leaves may shrivel and fall off, raising the pH level of the soil and hinders the ability to absorb essential elements such as iron, magnesium and zinc required for the production of chlorophyll in the leaves [10].

Biologists use various conventional techniques such as DNA and RNA hybridization techniques (southern and northern blotting), flow cytometric analysis, cloning and sequencing, immunofluorescent staining, reverse transcription polymerase chain reaction (RT-PCR), fluorescence in situ hybridization, western blotting, enzyme-linked immunosorbent assay (ELISA), marker-assisted selection, and gas chromatography with mass spectrometry to detect diseases in the citrus plants [12]. Phenol-chloroform, Chelex and solid-phase extraction techniques are used to hydrolyse and convert deoxyribose sugar to omega hydroxy acids and extract DNA from the citrus disease infected leaves [13]. The gel electrophoresis technique separates the macromolecules into fragments depending on the size and charge accumulated in these molecules [14, 15]. The PCR method creates multiple copies of the desired DNA segment, generating multiple duplicates and identifies the nucleic acid sequence made up of the four nitrogenous bases: adenine, thymine, cytosine, and guanine that are utilized for detecting citrus diseases. These conventional techniques are expensive and consume longer time to perform experimentation [14, 16]. Therefore, researchers are focusing on the advancement of modern light-based spectroscopic techniques. These techniques are non-destructive, label-free, and can serve as rapid screening tools capable of identifying citrus diseases during the early growth stage, and potentially differentiating the diseases with similar symptoms.

The optical characteristics of plant tissues can be determined by the way light interacts with them. Light interacts with the surface of leaves primarily through three distinct mechanisms: absorption, reflection, and transmission. This interaction can be used to estimate the types and amounts of various biochemicals present, as well as to identify structural and morphological changes in the citrus plants. Reflectance spectra of leaves characterize the biochemical composition of the pigments present on the leaves [17].

Typically, the visible spectrum (400-700 nm) detects the absorption of photons by chlorophyll, the near-infrared (NIR) spectrum (700-1100 nm) detects the absorption by dry matter, and the short-wave infrared (SWIR) region (1100-2500 nm) detects water absorption in leaves [18]. Leaves are primarily composed of four major elements (oxygen, carbon, hydrogen, and nitrogen) and the interaction of photons with chemical bonds leads to the combination of vibrations [19]. Plants absorb light energy primarily through chlorophyll located in specialized organelles called chloroplasts [20]. Chlorophyll captures light within the blue and red bands and reflects or allows light to pass through in the green band that results in the distinctive green colour of plants [21]. Leaf transmittance varies depending on factors such as leaf thickness, presence of pigments and internal structures. Transmittance enables light penetration into lower leaf layers and activates various physiological processes including photosynthesis. Scattering occurs when light interacts with various structures within plant tissues, such as cell walls and organelles [22]. Scattering of light within the canopy can change its direction, strength, and wavelength distribution, and offers essential insights into the biochemicals present in leaves.

Citrus diseases affect the healthy growth of citrus plants and lead to premature death. The optical spectroscopic techniques are modern techniques to identify change in different pigments such as chlorophylls, flavonoids, betacyanins, phycobilins, tannins, betalains, lignin, anthocyanins and carotenoids in leaves [23, 24]. Raman spectroscopy is one of the sophisticated vibrational spectroscopic techniques that can identify the chemical structure, polymorph, crystallinity and molecular interactions of pigments in the leaves [25]. Fluorescence spectroscopy is a highly sensitive analytical technique for detecting disease-induced changes in the amount of chlorophyll in citrus leaves and fruits [26]. It identifies the change in different fluorescent compounds (e.g., chlorophyll, carotenoids, and flavonoids) present in the leaves [27]. The IR spectroscopy is one of the most commonly used vibrational spectroscopy that measures the absorption of frequencies of molecules and characterizes the chemical structures or functional groups of the pigments of leaves within the NIR, SWIR and mid-IR ranges of electromagnetic spectra [28]. Hyperspectral cameras are the modern imaging-based spectroscopy equipment that captures multiple images within the spectral range of ultraviolet (UV), visible, NIR and SWIR wavelength ranges and identifies the change in biochemicals in citrus disease infected plants [29].

The utilization of artificial intelligence in optical spectroscopy can help in the timely detection of diseases in citrus plants. It has been used to develop predictive models that can estimate the severity and spread of the citrus disease [30]. In the past, the multiple types of neural networks within limitations have been utilized to classify leaves as healthy, symptomatic, or asymptomatic at an initial stage of citrus disease [31]. For example, the artificial neural network (ANN) algorithm establishes a non-linear connection between input data (spectra) and output parameters (leaf characteristics) that can be influenced by noise and fluctuations arising from environmental factors or measurement errors. The detection of citrus disease in its early stage can be challenging since symptoms may not be visible in the plants [32]. The concept of artificial intelligence is helpful in identifying citrus diseases, which might be difficult to detect through visual inspections even by trained scouts. The application of artificial intelligence in spectroscopy for citrus disease detection has potential to revolutionize the way of monitoring and managing the spread of citrus diseases.

The review outlines various label-free optical spectroscopic methods as screening tools for the early, effective, and reliable detection of citrus plant diseases in the context of precision agriculture. The objective is to gain insight into the detection of different citrus diseases in leaves by employing fluorescence, Raman, infrared (IR) spectroscopy and hyperspectral imaging spectroscopy methods. Furthermore, the recent applications of artificial intelligence in spectral analysis are reviewed as an effective statistical tool for classifying between healthy, symptomatic, and asymptomatic citrus plants based on morphology and chemometric characteristics. The review can further encourage the stakeholders in improving the advancement and commercialization of field-based optical tools to detect citrus disease in plants.

## 2 Spectroscopic techniques and methodologies

### 2.1 Fluorescence spectroscopy

Fluorescence spectroscopy is a modern analytical technique that is used to explain the structure and interactions of a molecule [33]. The principle of fluorescence spectroscopy involves exciting the sample using distinct wavelengths of light, and the energy is absorbed by the molecules, which enter into an excited energy state. Consequently, the molecule emits fluorescent light as it releases the absorbed energy and returns to its ground state. The spectrum is measured from the emitted light that provides structural and chemical information about the molecules. Changes in the emission wavelength of fluorescent light can indicate alterations in the molecular environment or structural rearrangements of the molecules. Fluorescence spectroscopy is a technique for identifying alterations in both the chemical and physical structure of fluorescent compounds found in citrus plants. By analyzing characteristic peaks in the fluorescence spectra, citrus diseases such as citrus canker, HLB, bacterial spot, CVC, Alternaria brown spot, and black pit can be identified [34]. The fluorescence spectra can also measure alterations in the levels of pigments and other metabolic compounds in response to citrus diseases. The red and far-red fluorescence can be used to assess changes in chlorophyll levels, while blue and green fluorescence can indicate variations in phenolic compounds within the epidermal cells and cell walls of green mesophyll cells. Furthermore, fluorescence peak ratios such as 450:650, 550:650, 450:750 and 650:750 can measure the growth, mineral deficiency, moisture stress and pathogen response between healthy and infected citrus plants.

A fluorescence spectroscope consists of four major components: an excitation source, optical fibers, optical filters and a spectrometer [26]. Light emitting diodes (LEDs) and lasers are used as excitation sources in the fluorescence spectroscopy. The laser sources of wavelengths of 365 nm, 405 nm, 470 nm and 530 nm with bandpass optical filters of 570 nm, 610 nm, 690 nm and 740 nm and LEDs of wavelengths 375 nm, 405 nm, 465 nm, 520 nm and 635 nm are often used to excite the plant tissues and record the fluorescence emission spectra [35-37]. The bifurcated probe delivers the light from the excitation source through a central fiber on the sample and collects back the emitted signals through multiple outermost fibers to the detector [35]. The sample is illuminated by the light from the excitation source after passing through a monochromator and the resulting fluorescence light is emitted in all directions. The detector receives a fraction of the fluorescent light that has passed through a secondary monochromator. The detector is typically a photomultiplier tube or filtered photodiodes [37]. Depending on the types of channels, there are two types of detectors: single-channel or multichannel detectors. The single-channel detector detects the intensity of a single wavelength. The multichannel detector can detect the intensities for the complete range of wavelengths in one snapshot and avoids the need of a monochromator. The dual monochromator equipped with a continuous excitation source is considered as the advanced fluorescence spectroscope capable of measuring emission spectra in relation to excitation spectra [38].

## 2.2 Raman spectroscopy

Raman scattering (or Raman effect) is an essential tool for quantitative and qualitative analysis of the biochemicals present in plant tissues. The polarity of a molecule affects the Raman scattering that simulates the vibration modes of the molecule and results in low-energy scattered photons. The anti-Stokes and Stokes lines denote shorter and longer Raman scattering wavelengths, respectively. Raman spectroscopy analyses chemical bonding and symmetry by examining their vibrational, and rotational, and low-frequency modes of the molecules [39]. The stokes Raman scattering occurs when the energy or frequency of the scattered photons is lower than that of the incident photon. The primary result of this phenomenon is elastic scattering (or Rayleigh scattering). The anti-Stokes Raman scattering takes place when the energy of the scattered photons exceeds that of the incident photon. Raman spectroscopy is employed to quantify larger molecules such as amino acids, carbohydrates, cellulose, lignin, lipids, and proteins in plants [40]. A change in the level of larger molecules can serve as an indicator for detecting infections in the leaves and fruits of citrus plants. Surface-enhanced Raman spectroscopy (SERS) is a modern approach for detecting stress-induced diseases caused by bacteria, fungi, and viruses in citrus plants [41]. The SERS is being considered as a better tool for the prevention and control of diseases at an early stage. The portable Raman spectrometers have less weight, are easy to operate, and consume less time for field analysis compared to other lab-based Raman spectrometers [42]. Microscopic imaging techniques are integrated with Raman spectroscopy and SERS to gain more insights into the change in the chemical structure of plant tissues.

Modern Raman spectroscopes use lasers as an excitation source within the UV, visible or NIR ranges [43]. The most commonly used lasers are 532 nm, 638 nm, 785 nm and 1064 nm [44]. The 532 nm laser is the second harmonic of the fundamental 1064 nm radiation generated from the neodymium-doped lasing material [45]. The 638 nm laser is a high-power single-mode diode that balances fluorescence with high signal resolution and is used for better acquisition of SERS data [46]. The 785 nm laser is relatively low cost in comparison to other lasers and has a good balance of detector efficiency, spectral resolution, limited interference from fluorescence background, and absorption of heat by the tissue sample [47]. The 1064 nm laser generates minimal fluorescence as the signal is much weaker while the spectra can be acquired with good SNR values [48]. The continuous-wave and pulsed-wave, both, lasers can be used in Raman spectroscopy. There are two types of detectors such as dispersive Raman and FT-Raman detectors that are used in Raman spectroscopy. The arrays of charge-coupled devices (CCD) are arranged in dispersive Raman detectors to optimize the detector for a wide range of spectra. The spectral range of dispersive Raman detectors can depend on the CCD detector size and the focal length of the spectrometer [48]. The FT-Raman detectors use NIR lasers and the detector is selected depending on the excitation wavelength. These detectors are manufactured from germanium, indium and gallium arsenide. Laser rejection filters such as notch and long-pass filters, serve the purpose of distinguishing between Raman and Rayleigh-scattered light. The subtractive mode of a triple-grating monochromator is used to isolate the desired signal. The holographic filter is used to record the very low Raman shifts [49].

## 2.3 Infrared spectroscopy

The IR spectroscopy is a widely used spectroscopic technique to study the interaction of IR radiation and matter through absorption, transmittance, or reflection modes [50]. There can be six types of vibrations involved in a molecule in interaction with the light such as symmetric and antisymmetric stretching and four other bending modes are scissoring, rocking, wagging and twisting of the molecule [51]. The absorptions take place at resonant frequencies when the absorbed frequency aligns with the vibrational frequency of the molecule [52]. The three primary factors influence the resonant energy such as (1) the configuration of the potential energy surfaces, (2) the masses of the atoms involved and (3) the vibronic coupling associated with the molecule. It identifies the presence of different functional groups from the characteristic bands of the respective intensity and frequency. The IR range could be majorly classified into three different ranges: NIR (700-2500 nm), mid-IR (2500-25000 nm) and far-IR (25000 nm-1mm) [53, 54]. The specific boundaries of the IR region may vary depending on the context and applications. In addition, the measurement of various pigments in leaves can be conducted across the visible-to-NIR range of spectrum (400-780 nm) [55]. The peak wavelengths for different chlorophyll pigments lie within the range of 430 to 696 nm [56]. The presence of anthocyanin and carotenoids are identified within the range of 530-550 nm and 420-503 nm, respectively [57]. NIR spectroscopy (NIRS)-based approach could be utilized as a rapid, non-destructive and, laboratory and field-based technique to identify healthy and infected plants grown in a field. The NIR reflectance spectral intensities at wavelengths 870 nm and 970 nm demonstrate higher accuracy in classifying healthy and HLB-infected orange leaves compared to the visible band at 570 nm and 670 nm [58]. In the mid-IR range, (Micro-)FTIR spectroscopy is utilized as a modern analytical technique to analyze spatially resolved chemical maps of plant tissues. This region is also sometimes referred to as the "IR fingerprint region" because several chemical compounds exhibit unique absorption and emission characteristics within this range [59]. For example, the peaks within the range of 5555-11111 nm have been utilized to identify changes in the amount of carbohydrate (starch) in orange leaves due to the infection of HLB disease. The far-IR region (also referred to as the terahertz -THz region) can detect freezing stresses in the citrus leaves [60]. It can identify the natural sensitivity to the vibrations of hydrogen bonds in aqueous solutions and these vibrations are generated with changes in the phenotype of plants [59].

Typically, citrus diseases can be detected from the change in the amount of different biochemicals in the leaves or fruits of the plants. However, in the scenarios when the disease-infected and nutrient-deficient plants might have similar phenotypes or patterns on the leaves and are often challenging to classify the two different classes of leaves [61]. Machine learning tools can utilize spectral data to distinguish between healthy leaves, symptomatic leaves, asymptomatic leaves, and leaves with nutrient deficiencies during the onset of the disease [62].

In the context of light sources, the deuterium and tungsten halogen excitation sources are often used in the UV, visible and NIRS [63]. A halogen gas is used in a tungsten lamp to enhance the effectiveness and luminosity of the light emitted. Halogen gases such as iodine or bromine react with the tungsten filament, allowing it to burn at a higher temperature and produce more light for the same amount of energy [64]. The lamps used with ground glass diffusion filter offer adequate illumination. These lamps are cost-effective and convenient to replace. They produce consistent visible light and dissipate excess energy as heat in the IR wavelengths [65]. A deuterium lamp is a gas-discharge lamp source that emits intense, continuous, and stable light within the wavelength range of 112-900 nm. This light is generated through the absorption and subsequent emission of ultraviolet radiation by deuterium gas [66]. The electric discharge formed within the bulb stimulates the molecular deuterium present in it and causes it to attain a heightened energy level. The deuterium releases light when it reverts to its original state and generates the continuous ultraviolet radiation [67].

The spectrometer consists of a dispersive element, collimator, an entrance slit, focusing optics, and a detector. The collimated light from plain grating then travels through a second spherical mirror that focuses the resulting diffracted light. The development of multi-element optical detectors such as CCD and complementary metal-oxide semiconductor (CMOS) detectors enables us to measure the fast scanning without the need for a moving grating [68]. The CCD detector is an integrated circuit consisting of a series of coupled capacitors that transfer the electric charge to the surrounding capacitor under the control of the external circuit [69]. The CMOS sensor is a type of image sensor that works on the principle of the photoelectric effect. It converts light into electrical signals and processes them to produce an image [70]. The CMOS sensors are typically less expensive, consume less power and can offer faster image readout speeds. The CMOS sensors have lower image quality compared to CCD sensors due to noise and lower dynamic range. A spectrograph interprets the spectrum into two main regions: functional group region (≥ 1500/cm) and fingerprint region (<1500/cm) [71].

### 2.4 Spectral analysis

In general, the acquired optical spectra are pre-processed to eliminate the noise that help to identify the spectral range having prominent spectral peaks. A dark and/or background spectrum is subtracted from the acquired raw spectra. Further, a baseline correction is performed to correct for any systematic offsets or slopes in the data [72, 73]. The normalization is performed to scale the spectral data to a common range or mean centre the data, which allows better comparison and analysis of spectral features. The standard normal variate (SNV) method normalization technique used to remove the effects of baseline shifts and intensity variations in spectral data [74]. The smoothing techniques such as Savitzky-Golay (SG) or moving average are applied to eliminate random spectral noise and elevate the signal-to-noise ratio. The SG filters attenuate high-frequency components and minimize the least-squares error when fitting a polynomial to noisy spectra frames. However, irrespective of the smoothing techniques applied, it is essential to reduce noise by distinguishing between the spectral bands to avoid the artifacts induced in the spectra [75]. Further next, the multiplicative scatter correction (MSC) method is applied to correct unwanted variations caused by scattering effects, instrumental changes, or sample inhomogeneity that effectively isolates and eliminates complex multiplicative effects that effectively isolates and eliminates complex multiplicative effects and, enhances the modelling of chemical effects that ultimately align the spectral shapes and improve comparability between different samples [76, 77].

### 2.5 Imaging spectroscopy approach

Hyperspectral imaging is a type of imaging-based spectroscopic technique that captures multiple images of the object across the visible, NIR, and SWIR regions. The principle behind this method is based on measuring the reflectance of an object across a continuous range of wavelengths within the chosen range of the electromagnetic spectrum, represented in the form of a hyperspectral cube [78]. The vertical and horizontal axes represent the pixel distances of the image and the third axis represents the wavelength. It is a modern form of imaging spectroscopy and is used in plant phenotyping [79]. Digital phenotyping is an advanced non-invasive sensing technique that automatically extracts morphological and physiological data from plants during phenotyping experiments [80]. It could be advantageous to utilize the technique for measuring the biochemical changes in addition to morphological and structural changes in disease infected or stressed citrus plants grown in a large farm [81]. Moreover, an aerial Light Detection and Ranging (LiDAR) system operating within a wavelength range of 250-2500 nm is utilized for detecting citrus diseases in farms [11, 82]. The combined technology of hyperspectral imaging and LiDAR is a modern remote sensing technology used to detect citrus diseases on large farms, especially in scenarios where manual inspection of every plant is not feasible.

Hyperspectral imaging system consists of tungsten-halogen lamps, objective lenses, imaging spectrograph and CCD or CMOS detector [83]. Halogen lamps are commonly used in indoor hyperspectral imaging experiments related to plants (Fig. 2). These lamps are made up of a transparent glass enclosure containing a tungsten filament with halogen gas and a small quantity of either iodine or bromine [64, 84]. These lamps provide a smooth and monotonic spectrum within the spectral range of visible (RGB camera) and NIR regions. The objective lenses determine the spatial resolution of the hyperspectral image. The selection of objective lenses is based upon the amount of incoming light from the observed area reaching the detector. The light captured by the objective lens then enters the spectrograph. The spectrograph consists of diffraction grating that disperses light into distinct wavelengths [82]. Most hyperspectral cameras available on the market use reflection gratings because of their beneficial characteristics that include the lack of high aberration, minimal distortion, low f-number, and broader field size [84]. The dispersed light is allowed to fall on the detector, where the photons are converted into electrical energy. The CCD and CMOS detectors are often used in hyperspectral imaging systems. Hyperspectral cameras capture the images within the extended wavelength range from UV, visible, NIR and SWIR ranges [83]. An airborne hyperspectral system consists of a hyperspectral camera, micro-computer system and navigation satellite system that are installed to capture the images within a continuous range of wavelengths of an area [79].

The acquisition of hyperspectral images involves obtaining three-dimensional hypercube data using four fundamental methods - point, line, area, and snap-shot scanning techniques [85]. The point scanning is a type of scanning technique that acquires the spectral information of the sample by scanning it one point at a time [86]. It consumes a significant amount of time for sample positioning and requires a stable experimental setup for high-quality image reconstruction. The line scanning technique is another type of scanning technique that acquires the spectral information of the sample by scanning it along a single line or several parallel lines together [87]. The area scanning is another type of scanning technique that scans two-dimensional monochromatic images at one wavelength at a time. The snap-shot scanning technique acquires hyperspectral images during an integration time of an array detector. The advantage of snapshot scanning is that it can acquire both spatial and spectral data within a shorter acquisition time and prevent motion artifacts. Image pre-processing is conducted following the completion of image acquisition [88]. The raw hyperspectral image is affected by the noise coming from the illuminating source, sensitivity of the detector and due to the transmission effect of multiple optical components in the system. The spectral calibration is necessary to eliminate these issues and is achieved through the use of white and black references. The black reference image is captured when both the light source and camera shutters are turned off. The white reference image is obtained by directing the camera towards a highly reflective white surface made of uniform materials (e.g., teflon and spectralon). The noise present in these hyperspectral images is majorly in the form of dead pixels or spikes [89]. The presence of dead pixels is a result of irregularities in the detector that result in permanent black pixels. The pretreatment techniques including spectral normalization, SG filtration, SNV and EMSC techniques are applied to eliminate the variations of the spectrum [90]. The derivative techniques are often applied to eliminate the constant offset or linear baseline shift. The segmentation divides the image into various distinct segments by considering their similarities in terms of spatial or spectral properties. Spectral segmentation requires a greater computational time when contrasted with spatial segmentation methods[89]. Following spectral pre-processing, the hyperspectral image data is typically corrected, transformed or enhanced in quality, subsequently utilized for feature extraction based on the objective (Fig. 3).

## 3 Applications of spectroscopic techniques in citrus diseases

Citrus plants are susceptible to several diseases infected by fungi, bacteria, viruses, and other pathogens that can impact their health and fruit production [91, 92]. The optical spectroscopic techniques use the interaction between electromagnetic radiation and matter to yield insights into the molecular composition of plant tissues [93]. These techniques have demonstrated numerous times the capability to identify changes induced by pathological infection in citrus plants, which are manifested through characteristic spectral patterns, thereby offering the potential of being a mass screening tool for rapid diagnosis in the field [25]. To harness the potential of these spectroscopic techniques, it could be implemented in conjunction with other methods such as integration to unmanned aerial vehicles or remote sensing approaches, which can further provide diverse applications to enhance the crop yields and prevent the transmission of diseases in the field [54]. This specific segment elucidates the various applications of these techniques and methods in the identification of numerous diseases in citrus plants, which are succinctly summarized in Table 1.

### 3.1 Applications of fluorescence spectroscopy

Fluorescence spectroscopy is a modern technique that can be employed in citrus orchards to study and identify specific disease symptoms, monitor disease progression and assess the effectiveness of disease management strategies [94]. The composition of multiple fluorophores in a leaf determines its emission spectra. The fluorophores identify changes in the chemical and physiological characteristics of citrus plants [36]. The major benefit of fluorescence spectroscopy is that it can be deployed as a real-time visualization tool in the agriculture fields [95].

Fluorescence spectral acquisition can be captured through a 10 mW green laser (532 nm) excitation that can stimulate lemon leaves and allow measurement of the chlorophyll fluorescence ratio between bands at 685 nm and 735 nm. This measurement helps distinguish between mechanical stress and disease stress caused by pathogens in citrus plants. Mechanical stresses experienced by plants are typically temporary and last for a short duration of time ranging from minutes to hours. However, a plant may take several days or even months for the symptoms of stress to manifest. Mechanical stress affects growth more slowly and is less severe than stress caused by pathogens [96]. The utilization of fluorescence spectroscopy allows for the differentiation of healthy, nutrient-deficient, and HLB-infected leaves in Valencia and Hamlin oranges. The yellow and green excitation fluorescence spectra are fitted into the bagging on decision trees classifier to classify the nutrient-deficient and HLB-infected leaves [27]. The ratio of fluorescence intensities of blue to red excitation (452:685) and blue to far-red (452:735) is used to detect the healthy and citrus canker infection in orange plants [94]. The blue-green and chlorophyll fluorescence were demonstrated to assess the various stages of infection in citrus canker disease [97]. In healthy and asymptomatic leaves, α-tocopherol and gallic acid exhibited higher fluorescence band intensities [98]. The fluorescence intensity was measured across a range of wavelengths from 450 nm to 800 nm in leaves using excitations at 405 nm and 470 nm. The selected bands were determined by changes in intensity caused by diseases and utilized to distinguish diseases with similar visual symptoms, such as citrus canker vs. citrus scab and HLB vs. zinc deficiency in citrus plants.

A violet-blue laser operating at 405 nm is utilised to record the chlorophyll spectra of grapefruit leaves infected by citrus canker. The 405 nm laser is often used as it lies within the 'soret' band of chlorophyll and can be used to study the structures and functions of botanical cells and tissues [35]. The 'soret' band appears as a wide absorption peak within 400-450 nm in the spectrum. It arises from the movement of electrons from the lower to the higher excited state [99]. The emission chlorophyll spectrum shows a major peak at 683 nm associated with the photosystem (PS) II whose primary function is to harvest light energy and use it to oxidize water molecules, releases oxygen and generates high-energy electrons that are passed on to other components of photosynthetic chain of electron transport [100]. Ultimately, the electron transport chain results in the generation of ATP and NADPH, which power the calvin cycle. The spectral band of 700-750 nm represents both PSs. PS-I consists of proteins complex, pigments, and cofactors that work together to capture light and transform it into chemical energy [101]. The core of PS-I contains a reaction center constituted by chlorophyll molecules that are capable of absorbing light in the red (around 700 nm) spectrum. These chlorophyll molecules absorb light and become excited and transfer their energy to a nearby molecule called a primary electron acceptor and initiates a chain of electron transfer reactions [102, 103] (Fig. 4).

The broad class of secondary metabolites known as phenolic compounds are found in the plants. The occurrence of different phenolic compounds is explained by the broad peak at 530 nm [35]. They are synthesized via the phenylpropanoid pathway and are present in different parts of plants. The most abundant and diverse classes of phenolic compounds include flavones, flavonols, flavanones, isoflavones and anthocyanins [104, 105]. Flavonoids, is another prominent category of phenolic compounds found in plants, encompassing hydroxycinnamic and hydroxybenzoic acids. As a disease progresses within the plant, the production of caffeic acid, tannins, flavins, chlorogenic acid including flavonoids are induced in the infected leaves of the plant [106]. These phenolic acids have demonstrated a variety of roles in biological functions including the antibacterial, antioxidant, and anti-inflammatory activities [107]. There are a group of phenolic compounds that include resveratrol, pterostilbene and viniferins. Stilbenes are another example of polyphenols, which is naturally produced by plants in response to various stressors such as UV radiation, fungal infections, insect harm or physical damage and, act as phytoalexins that help in protecting the plant from pathogens and environmental threats [108]. The most studied stilbene is trans-resveratrol, which has gained popularity for its potential health benefits. The fluorescence spectrum of trans-resveratrol typically exhibits an emission peak in the range of 320-340 when excited at a wavelength of around 280 nm [109]. Thus, the variations observed in fluorescence emission spectra can be utilized to gain valuable insights not only into specific plant metabolites but also into the multiple changes occurring in the local physiological environment, induced due to various external stresses and infections

### 3.2 Applications of Raman spectroscopy

Raman spectroscopy has emerged as a promising analytical technique for the characterization of citrus diseases in plants [48, 110]. This spectroscopic method utilizes the interaction between light and matter to detect the compositional and structural changes at molecular level occurring within the affected plant tissues to generate a unique spectral fingerprint of the sample under investigation. The technique enables the identification and characterization of specific biomolecules associated with citrus diseases, such as phytochemicals and metabolites [111]. This non-destructive and label-free technique offers several advantages, including quick analysis, high sensitivity, and the ability to probe samples in situ. It has demonstrated the ability to identify the unique spectral fingerprints of the pathogens and their byproducts in citrus plant tissues [112, 113]. The application of Raman spectroscopy in citrus disease research holds great potential for early and rapid disease detection, monitoring disease progression, and guiding targeted interventions for effective disease management in the citrus industry.

In 2019, Wang and their colleagues conducted a research wherein Raman spectra were acquired from intact leaves representing the class of asymptomatic, symptomatic, and healthy leaves. The measurements were conducted on the midrib section (mesophyll cells) of the leaves. Their study reported the capability of the Raman method to identify the three types of carbohydrates - glucose, sucrose, and starch present in leaves in different health conditions of the plant. The asymptomatic leaves exhibited relatively lower levels of carotene, sucrose, glucose, and chlorophyll in contrast to healthy leaves while an increase in the starch and polyphenol content were observed in asymptomatic conditions of the leaves. The BP-ANN model demonstrated a root-mean-square error (RMSE) of 0.0616 and a coefficient of determination ($R^2$) of 0.9598 [114] (Fig. 5).

Sanchez et al. conducted Raman analysis of nutrient-deficient, and HLB-infected leaves with healthy leaves of oranges and grapefruits. They observed that the Raman peak (1601−1630 $cm^{-1}$) associated with lignin, which significantly contributes to plant defence, was higher in severe-stage than the early-stage infected grapefruit leaves while no significant change was observed in case of orange leaves. In healthy grapefruit leaves, the xylan Raman peak (1218 $cm^{-1}$) intensity was observed higher compared to 1226 $cm^{-1}$, while the opposite trend was observed in HLB-infected grapefruit plants. Further, no discernible difference was observed between infected and healthy orange leaves. The Raman peak (1525 $cm^{-1}$) of carotenoids exhibited an increase in intensity of infected orange leaves. However, there was no change was observed in the case of grapefruit leaves. Polyphenols Raman peak (1247 $cm^{-1}$) was observed in nutrient-deficient grapefruit leaves while it was not prominently visible in orange leaves. These results led to the conclusion that these citrus plants exhibit cultivar-specific responses to different stresses. Moreover, it was also observed that the lignin Raman peak of nutrient-deficient leaves was significantly higher compared to healthy and HLB-infected leaves, which could be potentially utilised in differentiating the nutrient deficits from HLB infection [115].

HLB infected plants are vulnerable to the onset of citrus canker and blight that are other common diseases found in citrus plants. Early identification of various diseases, especially those with similar symptoms can play a crucial role in optimizing pesticide use and an effective management of the citrus diseases. The Raman spectra acquired from orange leaves revealed that the lignin and xylan spectral peaks were enhanced in HLB-infected leaves, whereas these intensities were reduced in leaves affected by citrus canker in comparison to the healthy leaves. The PCA analysis was performed in which the loading plot representing the peak of the lignin (1525 and 1630 $cm^{-1}$), cellulose (915, 1326, and 1630 $cm^{-1}$), hydrocarbon (1440, 1445, and 1448 $cm^{-1}$), xylan (1184 $cm^{-1}$), and carotenoid (1525 $cm^{-1}$) were utilised as predictors. The orthogonal PLS-DA enabled the differentiation of healthy plant from those affected by citrus canker, HLB, and HLB-blight, achieving an overall accuracy rate of 92% [25].

The combination of three techniques high-performance liquid chromatography (HPLC), Raman spectroscopy, and mass spectrometry can collectively enhance the accuracy in identifying the biochemical changes in citrus plants during HLB infection and the progression of the disease. The occurrence of HLB infection increases the content of p-coumaric acid and decreases the amount of lutein in the leaves. These changes were identified through Raman spectroscopy at an early stage. The combination of HPLC and Raman spectroscopy can identify the variations in carotenoid concentrations, which may provide an insight into the regulation of plants across various stages of their growth cultivated in both outdoor fields and controlled greenhouse environments [86, 116].

### 3.3 Applications of infrared spectroscopy

Vibrational spectroscopies, such as IR spectroscopy, can identify the presence of various biochemicals in citrus plants grown on large farms [117]. The IR spectral measurements can be acquired in three different modes such as absorption, reflection, or transmission of IR radiation by molecules present in leaves [118, 119]. It identifies the nature and structure of cell walls in plants. The spectra recorded within visible (400-750 nm), NIR (1100-2500 nm) and the mid-IR (2500-25000 nm) ranges can exhibit distinctive features that can be utilized to detect the disease or stress or deficiency in plants with a higher accuracy [120] (Fig. 6).

In three separate studies, the spectral properties of leaves infected with HLB in the NIR range were studied in which a field-portable spectroradiometer was used to measure the spectral data within 350-2500 nm. A portable halogen lamp (500 W) was used to excite the samples. The spectral features were analysed using different machine learning algorithms such as support vector machine (SVM) and k-nearest neighbour (k-NN) with an overall classification accuracy that varies in the range 80-97%. The results of all three investigations suggested that the visible-NIR region holds promise for distinguishing HLB-infected leaves. Moreover, it was emphasized that variations in different environmental factors under field conditions could lead to a decrease in accuracy. Hence, to enhance the accuracy in classification, it is essential to obtain multiple measurements of the same plant [121].

In another study, the evaluation of a four-band optical sensor consisting of two visible bands and two NIR bands, was conducted to identify HLB infection in citrus plants of Mid-sweet and Valencia orange cultivars. The four-band optical sensor classified the canopy of healthy and HLB-infected citrus plants at four selected wavelengths: 570 nm, 670 nm, 870 nm, and 970 nm. Using the five classification techniques (decision tree, k-NN, logarithmic regression, neural network, and SVM), the SVM achieved over 97% accuracy in distinguishing between HLB-infected and healthy plants using five measurements from each plant. The logarithmic regression exhibited the largest classification error among the five classification techniques exceeding 34%. The findings indicated that the four-band sensor has the potential to detect infected citrus plants at the symptomatic stage in field-based measurements. However, this study did not evaluate the capability of sensors to identify other stress factors including nutrient deficiencies. It might be possible that the sensor may provide more false positive results when exposed to other stressors. Furthermore, the performance of the sensor and classification algorithms applied on different citrus types could offer valuable insights and expand the uses of the tool [58].

Citrus rind disorders have significant impacts on citrus fruits that affect their appearance, quality, and marketability of fruits. This disorder does not manifest during harvesting or post-harvesting treatments but develops about 1-5 weeks after harvesting. The NIR-based methods can be utilized to determine the soluble solids content and total acidity in citrus fruits. Therefore, the NIRS based non-destructive tool could be explored to screen the citrus fruits based on their quality rather than damaged appearance caused by rind disorders [122]. Limited research has been conducted to develop effective technology capable of predicting and monitoring physiological symptoms in citrus rind disorders. The NIRS was employed to analyse the spectral properties of various fungal diseases found in citrus fruits, including citrus black spot, melanose, greasy spot, green mold, blue mold, sooty mold, and anthracnose. As this article primarily focuses on disease detection in citrus plants through their leaves, readers seeking information about fungal diseases, particularly in fruits, may refer to the cited published article [119].

The NIRS reflectance spectra acquired from dried orange and grapefruit leaves within the range of 400-2500 nm was studied. HLB-positive and HLB-negative leaves showed variations in visible range chlorophyll absorption peaks due to differences in the amounts of carbohydrates and cuticle waxes. The PLSR model classified these leaves with an accuracy of 92-99%. Laser-Induced Breakdown Spectroscopy (LIBS) is a powerful analytical technique that can be used for disease detection in citrus plants. The LIBS combined with NIR (i.e., LIBS-NIR) was utilized for the detection of HLB disease in citrus leaves. The analysis of atomic absorption spectroscopy and the Kolmogorov-Smirnov test indicated a reduction in calcium (Ca) and zinc (Zn) levels in the infected leaves. However, potassium (K) was found to be unaffected by the presence of the disease. The multi-layer perceptron PCA (MLP-PCA) achieved an accuracy of 89.5% during training and 95.7% during testing. It demonstrated the combination of LIBS and NIR is an efficient technique for identifying HLB disease in citrus leaves. [123].

Li et al. conducted a comparison of reflectance spectra between healthy and HLB-infected leaves. The elevated spectral intensities in visible bands attributed to the yellowish symptoms on infected leaves and reduced intensities in NIR bands show the impediments in water transportation to the leaves. The red edge position (REP) indicated a gradual shift towards the red bands with increasing infection levels. The threshold segmentation using REP achieved a classification accuracy of over 90%. The four-point linear extrapolation method proved more effective than the three-point Lagrangian interpolation method and showed the capability of REP as a valuable indicator for differentiating between healthy and HLB-infected leaves [124, 125].

By using the mid-IR spectroscopy, the processed HLB-infected and zinc-deficient leaves of Hamlin, Valencia, and Midsweet oranges in the form of powder were utilized for measuring spectral data in the ATR mode (Sankaran, Ehsani, et al., 2010). The change in the spectral peak within the range of 9000-10500 nm differentiated between healthy and infected leaves. This carbohydrate peak indicated the accumulation of starch in the HLB-infected leaves. The kNN algorithm achieved higher classification accuracy of over 90% than QDA algorithm within the wavelength range of 5150-10720 nm [54]. Utilizing the symptoms, even expert scouts often encounter challenges in distinguishing HLB and CVC diseases due to their similar visual symptoms (yellow blotches) on leaves. The spectral data from eight Valencia sweet orange leaves were collected using ATR-FTIR, and classified using PLSR algorithm. The results showed an accuracy of 93.8% in identifying healthy, CVC-symptomatic, and HLB-infected (both symptomatic and asymptomatic) leaves based on the presence of carbohydrates and other secondary metabolites [7].

Hawkins et. al., measured data from dried HLB infected orange and grapefruit leaves, which was grinded into powder form. The spectra were acquired using FTIR-ATR spectroscopy within 5882-11111 nm. The carbohydrates peaks within the 8474-11111 nm were utilized in identifying infected plants and non-infected plants with an accuracy greater than 95% using the chemometrics model (Hawkins et al., 2010). In another study by Hawkins et. al., the FTIR spectra of HLB and other citrus maladies were compared. The spectra of processed grapefruit leaves were measured using the FTIR spectrometer with a deuterated triglycine sulfate (DTGS) detector within 5882-14285 nm. The spectra of HLB-infected leaves showed a sharp and asymmetrical peak at 9803 nm and two sharp peaks at 8695 nm and 9285 nm while healthy leaves showed a broad carbohydrate flat peak centred at around 9505 nm and two weak peaks at 8695 nm and 9285 nm. The PCA and multiple linear regression were analyzed to classify the HLB, nutrient deficiencies, and citrus diseases in leaves. The study demonstrated that different citrus diseases and deficiencies exhibit similar carbohydrate transformations as observed in the spectra of HLB-infected plants [28] (Fig. 7).

To explore changes in carbohydrates (starch and soluble sugars) within the phloem of citrus leaf midribs; the study by Yang et. al., reported a semi-quantification model to predict carbohydrate concentrations within 4000–675 $cm^{-1}$ using the micro-FTIR spectroscopy. The least squares support vector machine regression (LS-SVR) model in conjunction with the RF algorithm achieved the most accurate predictions with determination coefficients of prediction ($R^2P$) values reaching 0.85 and 0.87 for starch and soluble sugar concentrations in leaf midribs. Moreover, the multi-layer perceptron classification models demonstrated a classification accuracy of 94% and 87% in identifying HLB disease using the complete spectral range and the optimal wavenumbers determined through the RF algorithm, respectively. The technique has the potential to be an essential tool for predicting carbohydrate concentration and detecting HLB disease [59].

### 3.4 Application of imaging spectroscopy

Hyperspectral imaging is a modern method that analyzes the composition and characteristics of plant tissues by capturing and examining their spectral data. It is a powerful optical sensing technique used to collect and analyze spectral data typically captured in the range of visible light (400-700 nm) and extends into the NIR and SWIR regions, which can range from around 700 to 2500 nm [126, 127]. Hyperspectral imaging can identify disease symptoms at early stages before they become visually apparent. It quantifies the severity of citrus diseases by analyzing the extent of spectral changes in infected plants [125, 128]. It has emerged as an indispensable tool in citrus disease management, facilitating early detection, precise classification, and monitoring of disease progression [129].

Indoor (lab-based) and outdoor (portable) hyperspectral imaging systems were utilized for the detection of citrus disease in orange, lemon, lime, and grapefruit plants. Hyperspectral images of citrus leaves were acquired at varying spectral ranges depending on the instruments used. Wang et. al., acquired hyperspectral images from healthy and HLB-infected navel orange plants within 400-1000 nm. A method was proposed for the instantaneous detection of symptomatic and asymptomatic leaves to rapidly identify citrus HLB disease. The comparative analysis of five different data pretreatment methods revealed that the least squares-SVM (LS-SVM) based on adaxial and abaxial leaf surface spectra provide the best recognition results for classifying healthy and infected orange leaves. It achieved training and testing accuracies of 100% and 92.5%, respectively [114]. Li et al. utilized hyperspectral imaging along with the competitive adaptive reweighted sampling (CARS) and LS-SVM algorithms to measure chlorophyll present in lemon leaves infected with citrus yellow vein clearing virus (CYVCV). During the onset of CYVCV, hyperspectral images of lemon leaves infected with CYVCV were acquired within 300-1100 nm and revealed a reduction in the amount of chlorophyll in mesophyll tissues near the veins of affected leaves. The LS-SVM models were used to predict chlorophyll content in lemon leaves. The characteristic wavelength peaks at 550 nm, 680 nm and 680–750 nm were selected to identify healthy and CYVCV infected leaves. The prediction model utilized wavelengths extracted through the CARS algorithm achieved a RMSE of 0.10, a relative prediction deviation of 3.91, and a determination coefficient of 0.94 in the testing set. The observed results indicated that the method could help in rapid detection and quick visualization of chlorophyll levels and in identifying the symptoms of CYVCV disease in lemon leaves [130].

The combination of hyperspectral imaging and carbohydrate metabolism analysis was proposed to create whole-season classification models for detecting citrus HLB disease. This approach is particularly useful due to the extended asymptomatic period of HLB infection and the similarities in symptoms with nutrient-deficient citrus plants. The accumulation of sucrose in the infected leaves was observed steadier compared to fructose, glucose, and starch throughout both the hot and cool seasons while an opposite pattern observed in iron deficient leaves. The LS-SVM technique achieved classification accuracies of 90.2% for healthy, 96% for HLB-infected, and 92.6% for iron-deficient leaves within 450-1023 nm across different seasons. [131].

Among remote sensing applications, airborne hyperspectral imaging approaches have demonstrated their suitability at the farm and regional scales. Unmanned aerial vehicle (UAV) technologies offer a promising avenue for on-field and farm surveillance, effectively curbing data acquisition expenses. While fewer studies focused on a regional scale for plant disease evaluation, those investigations demonstrate the potential for upscaling remote sensing-based approaches. A point of particular interest in integrating hyperspectral imaging tools with UAV technologies as an effective intervention for differentiation of disease infections and monitoring their stages of development in the farm field [129].

Li et. al. conducted an experiment employing the potential of an airborne hyperspectral imaging system to acquire images for detecting HLB disease. Spectral angle mapping using pure endmembers helped differentiate healthy and infected areas with HLB disease. The experimental outcomes were compared using supervised and unsupervised methods, yielding accuracies within the range of 40.9% to 63.6%. The focus of this study was the evaluation of the proposed novel approach termed as 'extended spectral angle mapping (ESAM)', which was introduced for the detection of HLB disease. The algorithms calculate the spectral similarity between two spectra of each pixel in an image in terms of the angle between them. ESAM exhibited superior performance, achieving a detection accuracy of 86% compared to those other two methods. The findings of the study illustrated that the accuracy of detection in airborne hyperspectral imagery could be elevated by implementing the pure endmember extraction and inclusion of the red-edge position of HLB and thereby demonstrated the potential for the detection of HLB disease using hyperspectral imagery [125]. Furthermore, an improvement was made by Xiaoling et al. in which another model, namely the fully connected neural network, was employed. The research investigated combining various spectral features to improve the model potential to identify HLB-related features and enhance its resistance to interference. The approach provided a more intuitive way of capturing the disease distribution within the canopy, offering valuable insights for the precise removal of branches and stems affected by HLB [132].

Francisco et al. conducted a study wherein hyperspectral images of Valencia orange plants captured by both aircraft and UAVs within the 530-900 nm range were classified into healthy and HLB-infected plants. The results obtained using the UAV-based hyperspectral imaging sensors have a higher resolution (5.45 cm/pixel) compared to aircraft-based sensors which have a lower spatial resolution (0.5 m/pixel). The stepwise regression analysis was used to identify relevant features from spectral images captured by both UAVs and aircraft. The differences were observed in the 710 nm and NIR-R index can classify between healthy and infected plants at both spatial resolutions. The utilization of a kernel-enhanced SVM outperformed in classifying hyperspectral images of healthy and infected orange plants. The UAV-based system demonstrated a classification accuracy range of 67-85% while aircraft-based sensors showed an accuracy range of 61-74%. Therefore, in cases of symptoms at the branch level images taken with high-resolution aerial sensing techniques (UAV) at low altitudes demonstrated promising results for the detection of infected plants. Scouting is a key practice in HLB disease control, the study indicates that this approach could serve as a quick sensing technology to assist scouts, thereby reducing scouting costs and enhancing scouting efficiency [134, 135] (Fig. 8).

The exploration of an integrated approach involving hyperspectral imaging, fluorescence, and thermal imaging for crop monitoring remains largely unexplored, having only been initiated. The combination of an airborne system encompassing optical hyperspectral, thermal, and fluorescence data carries the promise of early detection capabilities in the field and facilitating improved monitoring of plant diseases. The success of such integration has already been demonstrated by studies focusing on quality control and fruit safety [11].

**Table 1** Summary of optical spectroscopic techniques for detection of citrus disease in leaves.

| References | Samples | Spectral bands | Spectroscopic techniques | Data analysis methods | Major work / observation |
|---|---|---|---|---|---|
| **[136]** | Orange leaves | 685-735 nm | Fluorescence spectroscopy | Figure of merit | Fluorescence spectral analysis was conducted to discriminate citrus canker infected leaves from healthy leaves (excitation lasers: Nd:YAG-532 nm and HeCd-442 nm). |
| **[96]** | Rangpur orange leaves | 680–712 nm and 712–750 nm | Fluorescence spectroscopy | Figure of merit | Two chlorophyll fluorescence bands ratio was used to detect and discriminate between mechanical stress and citrus canker induced-stress in plants (Nd:YAG excitation laser -532 nm). |
| **[128]** | Grapefruits | 553 nm, 677 nm, 718 nm, and 858 nm | Hyperspectral imaging | PCA | Classification of hyperspectral images to distinguish citrus canker from healthy fruits and other diseases, including wind scar, cake melanose, greasy spot, specular melanose, and copper burn. |
| **[126]** | Grapefruits | 450-930 nm | Hyperspectral imaging | Spectral information divergence (SID) mapping | Classification of hyperspectral images of citrus canker from healthy fruits and other diseased peel conditions such as melanose, wind scar, scab, insect damage, and greasy spot. |
| **[97]** | Citrus leaves | 547–620 nm, 680–712 nm, 712–750 nm, and 680–800 nm | Fluorescence spectroscopy | Figure of merit | Distinguishing healthy leaves from citrus cankered leaves using fluorescence spectra obtained under laboratory conditions with an Nd:YAG excitation laser (532 nm). |
| **[94]** | Mandarin orange, sweet orange, key lime, Persian lime and Rangpur lime leaves | 520-540 nm, 680–712 nm, 712-750 nm and 680-800 nm | Fluorescence spectroscopy | Figure of merit | Laser induced fluorescence spectra (excitation: Nd:YAG-532 nm) were analyzed to distinguish citrus canker from CVC, citrus scab, and HLB diseases. The diagnostic method for citrus canker |
| | | | | | exhibited high sensitivity (~90%) but, low specificity (~70%), possibly due to signal collection from a small area of the leaf. |
| **[137]** | Orange and grapefruit leaves | 900-1150 $cm^{-1}$ | ATR-FTIR spectroscopy | Principal component regression and PLSR | The ATR-FTIR spectra of leaves showed a rise in the carbohydrate peak within the range of 900–1180 $cm^{-1}$, which was used to distinguish between healthy and HLB-infected plants. |
| **[54]** | Hamlin orange, Valencia orange and Midsweet orange leaves (powder) | 952–1112 $cm^{-1}$ | Mid-IR spectroscopy | QDA and k-NN | The carbohydrate peak within the range of 952–1112 $cm^{-1}$ showed the accumulation of starch in HLB-infected leaves. This spectral peak differentiated the infected leaves from healthy and nutrient-deficient leaves. |
| **[121]** | Orange leaves | 350–1000 nm, 1000–1850 nm and 1850–2500 nm | NIR spectroscopy | QDA and SIMCA | Identification of spectral features within 350-2500 nm to classify the NIR spectra of healthy leaves from HLB infected leaves using the machine learning techniques. |
| **[58]** | Valencia orange leaves | 570 nm, 670 nm, 870 nm, and 970 nm | NIR spectroscopy | k-NN, SVM and decision trees | HLB detection in citrus leaves was conducted using four narrow-band sensors, and the spectra were classified using the machine learning techniques. |
| **[27]** | Hamlin and Valencia orange leaves | - | Fluorescence spectroscopy | Bagged decision tree and Naïve Bayes classifier | Healthy and HLB-infected leaves were differentiated by analysing yellow, red, and far-red (NIR) fluorescence spectra when excited with UV, blue, green, and red wavelengths under both laboratory and field conditions |
| **[7]** | Valencia sweet orange leaves (powder) | 900-1175 $cm^{-1}$ | ATR-FTIR spectroscopy | PLSR | Healthy, HLB, and CVC-symptomatic leaves were identified based on changes in the amount of secondary metabolites |
| | | | | | (hesperidin and umbelliferone) and carbohydrates, using the ATR-FTIR spectra. |
| **[95]** | Navel oranges, Pineapple orange, Sunburst, Murcott, Rhode Red Valencia, Hamlin, Valencia, Grapefruit leaves. | 590 nm, 688 nm, and 750 nm. | Fluorescence spectroscopy | Bagged decision tree and Naïve Bayes classifier | Yellow, yellow-to-far-red, and far-red fluorescence spectra were measured from healthy, HLB-symptomatic and HLB-asymptomatic leaves. Healthy and HLB-symptomatic leaves were classified with an accuracy of 97% using a bagged decision tree classifier (excitation wavelengths: ultraviolet-375 nm, blue-465 nm, green-520 nm, and red-635 nm). |
| **[134]** | Citrus orchard | 530 nm, 560 nm, 660 nm, 690 nm, 710 nm, and 900 nm | Hyperspectral imaging | SVM | High-resolution aerial imaging of a citrus orchard was captured using a hyperspectral camera mounted on a remote sensing multi-rotor UAV to classify the healthy and HLB-infected citrus trees using the SVM technique. |
| **[125]** | Citrus leaves | 700 nm | Hyperspectral imaging | Extended spectral angle mapping (ESAM) and K-means | Detection of healthy and HLB-infected plant canopies from airborne hyperspectral images and these images were classified using different machine learning techniques. |
| **[138]** | Citrus leaves | 229.7 nm, 247.9 nm, 280.3 nm, 393.5 nm, 397.0 nm, and 769.8 nm | Laser-induced breakdown spectroscopy | QDA and SVM | Classification between healthy leaves and other similar anomalies such as citrus cankered, HLB-infected, and nutrient-deficient leaves with an average accuracy of 97.5% using the SVM technique. |
| **[36]** | Mandarin orange, sweet orange, key lime, Persian lime and Rangpur lime leaves | 550 nm, 560 nm, 570 nm, 580 nm and 690nm | Fluorescence imaging spectroscopy | SVM | Classification of HLB-infected from zinc deficient leaves and citrus canker from scab using fluorescence images with higher accuracies of 95% and 97.8%, respectively. |
| **[112]** | Sweet orange, Persian lime and Mexican lime leaves | 905 $cm^{-1}$, 1043 $cm^{-1}$, 1127 $cm^{-1}$, 1208 $cm^{-1}$, 1370 $cm^{-1}$, 1272 $cm^{-1}$, 1340 $cm^{-1}$, and 1260–1280 $cm^{-1}$ | Raman Spectroscopy | PCA-LDA | Raman spectra were analysed to discriminate between healthy and HLB-infected leaves. The spectral signature related to the presence of amino acids, proteins, carbohydrates, and lipids were identified. PCA-LDA analysis showed a precision of 89.2 % in differentiating HLB-infected leaves from healthy leaves. |
| **[23]** | Citrus plant | 960 $cm^{-1}$, 1087 $cm^{-1}$, 1109 $cm^{-1}$, 1154 $cm^{-1}$, 1225 $cm^{-1}$, 1385 $cm^{-1}$, 1462 $cm^{-1}$, 1707$cm^{-1}$, 2882 $cm^{-1}$, 2982 $cm^{-1}$, and 3650 $cm^{-1}$ | FTIR-ATR spectroscopy | Spectral peak analysis | Pentanone was identified as a spectral marker for distinguishing between healthy and HLB-infected plants caused by Ca. L. The marker was observed as absent in the infected plants. |
| **[139]** | Sweet orange leaves | 410-630 nm and 650-800 nm | Fluorescence spectroscopy | PLSR | Healthy, HLB-infected, and CVC-infected leaves were detected using fluorescence spectra acquired by a 405 nm excitation laser. Asymptomatic leaves were identified 21 weeks before symptoms appeared on the leaves with an accuracy exceeding 90%. |
| **[140]** | Sweet orange leaves | 279.55-285.21 nm, 315.89-317.93 nm, and 766.49-769.90 nm | Laser-Induced Breakdown Spectroscopy | PCA and PLSR | Classification between healthy, HLB-asymptomatic, and HLB-symptomatic leaves was achieved with an accuracy of 73% using the PLSR technique. |
| **[141]** | Citrus leaves | 620 nm | Fluorescence imaging spectroscopy | Random frog, sequential forward selection, and Monte Carlo uninformative variable elimination methods | Fluorescence images of healthy, HLB-infected, and nutrient deficient leaves were discriminated with an accuracy of 97% based on the combination of features and mean fluorescence. |
| **[142]** | Citrus plants (phloem tissues) | 200-700 nm | Laser-induced breakdown spectroscopy | PCA | Classification accuracy of 88% or higher between healthy and HLB-affected plant phloem was achieved by applying PCA analysis to laser-induced breakdown spectra. The differences in the healthy and infected spectra were attributed to the characteristic lines corresponding to elements such as calcium, sodium, nitrogen, hydrogen, and iron present in the phloem samples (excitation: Nd:YAG 1064 nm pulsed laser). |
| **[131]** | Satsuma leaves | 493 nm, 515 nm, 665 nm, 716 nm, and 739 nm | Hyperspectral imaging | LS-SVM | Three-class classification models and the LS-SVM classifier were used to classify between healthy, HLB-infected (asymptomatic and symptomatic), and nutrient-deficient leaves with accuracies of 90.2%, 96.0%, and 92.6% for the cool season, hot season, and the entire period, respectively. |
| **[98]** | Grapefruit and Mexican lime leaves | 466 nm, 515/536 nm, 680 nm, 730 nm and 780 nm | Synchronous fluorescence spectroscopy | PCA | The decrease in the chlorophyll bands and increase in blue-green region indicated the severity of citrus canker disease in leaves. Fluorescence spectra of healthy and asymptomatic leaves showed a high intensity of gallic acid and α-tocopherol, while symptomatic leaves exhibited a high intensity of caffeic acid, flavonoids, flavins, tannins, and chlorogenic acid. |
| **[11]** | Sugar Belle leaves | 700-950 nm | Hyperspectral imaging | k-NN | The classification of citrus canker in asymptomatic, early symptomatic, and late symptomatic leaves showed accuracies of 94%, 96%, and 100%, respectively. |
| **[110]** | Orange and grapefruit leaves | 1180 $cm^{-1}$, 1155 $cm^{-1}$, 1329 $cm^{-1}$, 1455 $cm^{-1}$, 1528 $cm^{-1}$, and 1604 $cm^{-1}$ | Raman spectroscopy | Orthogonal PLS discriminant analysis | Raman spectra acquired using an 831 nm excitation laser and a 2mm beam size, have demonstrated detection rates of approximately 98% for grapefruit and approximately 87% for oranges when distinguishing between healthy, nutrient-deficient, and HLB-infected plants. |
| **[114]** | Navel orange leaves | 614–1722 $cm^{-1}$ (730–810 $cm^{-1}$, 866 $cm^{-1}$, 942 $cm^{-1}$, 1082 $cm^{-1}$, 1250 $cm^{-1}$, 1455 $cm^{-1}$, and 1510–1630 $cm^{-1}$) | Raman spectroscopy | PCA, PLS-DA, and BP-ANN | The study demonstrated the capability of the Raman method to identify the three types of carbohydrates - glucose, sucrose, and starch present in leaves in different health conditions of the HLB-infected plant. The asymptomatic leaves exhibited relatively lower levels of carotene, sucrose, glucose, and chlorophyll in contrast to healthy leaves while an increase in the starch and polyphenol content were observed in asymptomatic conditions of the leaves |
| **[113]** | Orange and grapefruit leaves | 942-944 nm | Raman spectroscopy | Orthogonal PLS discriminant analysis | The findings of the work indicated that Raman spectroscopy (laser excitation 830 nm) has the potential for significantly more sensitive detection of HLB when compared to quantitative polymerase chain reaction. |
| **[35]** | Grapefruit leaves | 530 nm, 686 nm and 735 nm | Fluorescence spectroscopy | PCA and PLSR | Detection of citrus canker in leaves based on the concentration of phenolic compounds, such as alkaloid berberine and quercetin, and observed chlorophyll using the fluorescence spectra. |
| **[129]** | Citrus plants (orchard) | 468 nm, 504 nm, 512 nm, 516 nm, 528 nm, 536 nm, 632 nm, 680 nm, 688 nm, and 852 nm | Hyperspectral imaging | Quadratic SVM | UAV hyperspectral imaging was used to detect HLB disease on large-scale citrus orchard plants and images were classified using a stacked autoencoder neural network with an accuracy of 99.33%. |
| **[123]** | Gannan navel orange leaves | 5100 $cm^{-1}$ and 6800 $cm^{-1}$ | LIBS-NIR spectroscopy | DA-PCA and MLP-PCA | The combination of LIBS and NIR achieved the highest diagnostic accuracy of 95.7% in detecting HLB. The findings indicated a significant decrease in the levels of minerals like zinc (Zn), magnesium (Mg), and calcium (Ca) in HLB-infected leaves, resulting in the appearance of yellow areas and spots. The LIBS–NIR joint method was proposed as a better analytical technique for identifying HLB in compared to using either LIBS or NIR individually. |
| **[111]** | Grapefruit leaves | 400 $cm^{-1}$ - 2000 $cm^{-1}$ (1525 $cm^{-1}$, 1575 $cm^{-1}$, 1601 $cm^{-1}$ and 1630 $cm^{-1}$) | Raman spectroscopy | Kruskal-Wallis one-way analysis of variance | The analysis of Raman spectra indicated an increase in p-coumaric acid (1601 $cm^{-1}$ and 1630 $cm^{-1}$) content while a decrease in lutein (1525 $cm^{-1}$) content in the HLB-infected leaves. The results imply that Ca. L. asiaticus triggers a defence response designed to eliminate the bacteria within the phloem. |
| **[59]** | Orange leaves | 1175–900 $cm^{-1}$, 1500–1175 $cm^{-1}$, and 1800–1500 $cm^{-1}$ | Micro-FTIR spectroscopy | LS-SVR and RF | Micro-FTIR spectroscopy was utilized to estimate carbohydrate concentrations in the midribs of healthy, HLB-infected (symptomatic and asymptomatic), and nutrient-deficient plants using three fingerprint regions, specifically 1175–900 $cm^{-1}$, 1500–1175 $cm^{-1}$, and 1800–1500 $cm^{-1}$. The spectral intensities of these peaks, as measured in HLB-asymptomatic and HLB-symptomatic leaves, were observed higher in comparison to the healthy and nutrient-deficient plants. |
| **[130]** | Lemon leaves | 305nm –1090 nm (550 nm, 680 nm and 680-750 nm) | Hyperspectral imaging | CARS and LS-SVM | The findings of the work indicated that hyperspectral imaging in the range of 305-1090 nm combined with CARS and LS-SVM algorithm, offers an efficient way to assess the distribution of chlorophyll in CYVCV-infected lemon leaves, providing an enhanced understanding of the disease symptoms. |

## 4. Limitations and Future Prospects

The plant leaves have complex structures with multiple layers of cells, pigments, and other compounds. The absorption, scattering and penetration depth of light into the plant leaves might vary due to variations in the thickness of leaves, which consequently limits the technique to the measurement of surface properties of plant leaves. Thus, the technique might not represent the physiological status of the entire thickness of the leaf. However, the acquired surface properties might be sufficient in detecting the chemical or structural changes in leaves, which is not visible by eyes and might require a laboratory-based test. Although recent studies have indicated that optical spectroscopy is rapid, label-free, non-destructive, and non-invasive technique that can be utilized directly in the field as screening tool to detect diseases in citrus plants, there are several challenges that exist in the commercialization of spectroscopic techniques. The environmental noises and high light intensities can affect the measurement of the spectra. It is often needed to select an appropriate wavelength or index especially in case of NIRS that is not affected by the surrounding noises in the environment. A single spectral measurement might not be sufficient for accurate detection of infected plants; therefore, multiple measurements from the same plant can enhance detection and classification accuracy. Furthermore, conducting multiple measurements can additionally reduce the impact of daylight and other variations in ambient conditions (such as temperature, humidity, and air quality) on optical spectra, if present. A field-based portable or hand-held spectrometers might generate thermal noise and consequently degrade the quality of the spectra while operating continuously for a long time. Certain advanced equipment, particularly those which are still in the experimental stages, may necessitate a specialized or expensive instrumental setup. However, with advancements in optical and detector technologies, the incorporation of more machine learning-based algorithms or artificial intelligence-based software, and an increase in the number of users, the cost reduction is highly expected in near future.

Among the array of nondestructive and rapid assessments techniques and tools employed for quality assessment, NIR spectroscopy stands as a relatively more established option in terms of its instrumentation, compatible accessories and its suitability with readily available chemometric packages. NIRS exhibits a fairly high degree of ruggedness. However, the robustness of the technique relies on several factors including the sensitivity of the sensor employed. Although numerous investigations have demonstrated promising results when utilizing the NIR spectroscopy method to detect and discriminate diseases in plants, it is imperative to exercise prudence in analysis of associated spectral data. This is due to the fact that absorption bands in NIR spectral data are typically broad and are often subject to interference resulting from overlapping absorptions. This interference primarily attributed to overtones and combinations of vibrational modes involved in various chemical bonds such as N–H, O–H, C–H, and S–H. Another major challenge encountered in NIR spectroscopy-based method is due to the intrinsic variability in plant leaves. Consequently, disparate NIR configurations and analytical frameworks have been employed in pursuit of enhanced predictive capabilities.

However, it is essential to acknowledge that the selection of these parameters can exert a tangible influence on both the accuracy and robustness of the predictive models. Therefore, it is of one of the major challenges in the NIRS is to select, optimize and standardize the settings and analytical framework to acquire best results with higher accuracy and make it more viable for agri-commercial application. Another issue associated with NIRS technology involves the necessity for a calibration step for every new set of measurements. The challenge could be mitigated by automating this process using a standard material. The approach has begun to be incorporated into portable and handheld devices in manual measurements. However, implementation of such approach with UAV based measurements might be challenging. Creating and maintaining extensive spectral libraries encompassing various citrus plant species under diverse conditions and geographical locations, demands substantial resources and time. The existence of such spectral libraries can enhance the practicality of NIRS models in precision agriculture.

Fluorescence spectroscopy is another major optical tool in plant spectroscopy, offering insights into the molecular composition and physiological status of plants. However, the intricate structures of plant tissues, encompassing nonuniform cell walls, pigments, and heterogeneous fluorophore distribution present challenges to the technique that can compromise the precision and interpretation of fluorescence signals. Environmental factors such as ambient temperature, light intensity, and humidity can also influence fluorescence signals, making it challenging to obtain consistent results across different experimental conditions. Photobleaching is one of the major limitations of the method in which exposure of laser light leads to the degradation of the fluorophore and results in the loss of its ability to fluoresce and, consequently the loss in fluorescence signal is encountered. It reduces the sensitivity and accuracy in recording of the spectrum from the leaf sample. Photoquenching might also cause the reduction in the fluorescence intensity of fluorescent molecules under high-intensity light exposure, but at the same time it also offers an additional viable approach for investigating the dynamics of endogenous fluorophores and their interactions within the local environment of plant tissue. The photobleaching may be overcome by adjusting the intensity and duration of exposure of light and selecting longer excitation wavelengths whenever possible. While longer excitation wavelengths can reduce photobleaching, it might also contribute in less efficient excitation of certain fluorophores. Therefore, the selection of excitation wavelength needs a balance between minimizing photobleaching and effectively exciting the target fluorophores of interest. Different citrus plants and species might have varying fluorophores and susceptibilities to photoquenching and photobleaching, necessitating customized optimization of methods and protocols.

Raman spectroscopy offers numerous advantages when compared to other spectroscopic techniques specially for measurement of plant leaves. It has the capability to accurately identify bio-compositional changes by attributing distinct fingerprint characteristic peaks, which aids in uncovering the underlying mechanisms associated with the detection and progression of plant disease. In general, Raman scattering cross-section is relatively low that can limit the sensitivity of the technique for the analytes with low concentrations available in the sample volume. Raman spectroscopy is susceptible to endogenous fluorescence present in plant leaves and can interfere with the Raman signal. Sometimes the fluorescence caused by impurities present in the sample or by the excitation wavelength can mask the Raman signals. The spectral resolution of Raman spectroscopy is limited by the spectral width of the laser source used and might makes it difficult to resolve closely spaced spectral features. Moreover, the prolonged exposure of the high-power laser during data acquisition on the plant samples may damage or alter the compositions of the plant tissues. One of the primary challenges in citrus diseases lies in distinguishing between biotic factors (typically caused by pathogenic infections) and abiotic stressors (usually resulting from unfavourable environmental conditions, including nutrient deficiencies) because these two factors might manifest as similar visible symptoms. Raman-based technique has high potential to differentiate such stressors in plants, especially when applied in the field conditions. The alterations in metabolic factors indicate that plants responding to both biotic and abiotic stresses utilize distinct regulatory mechanisms to enhance their resistance to stress. Therefore, conducting specific assessments of the diverse chemical compositions, which contribute to inducing a specific phenotype in citrus plants, can significantly enhance the practical utility of the technique in the near future.

Hyperspectral imaging is an expensive and powerful imaging technique that can capture the detailed spectral information about the composition and properties of objects. The images encompass a large amount data and require significant processing power and storage space to handle. This makes it difficult to analyze large datasets in real-time. Typically, a standard hyperspectral image often sacrifices spatial resolution for spectral information. A lower spatial resolution in comparison to conventional color image can limit the ability to resolve small features in a field-of-view of a plant. The image can be sensitive to environmental or atmospheric conditions, which can affect the quality of the data. Another challenge is that hyperspectral imaging typically has a limited field-of-view, which can make it challenging to capture data over a large area. The spectral response of different plants can vary depending on the number of factors such as lighting conditions or surface orientation. This can make it challenging to accurately identify or classify the plants based on their spectral signature. Additionally, there is a need for systematic comparisons between leaf-scale and canopy-scale detection methods and the severity of citrus disease infestations using the technique of hyperspectral imaging. The inclusion of both types of data can enhance the variability of the dataset, the robustness of the model, and consequently, the usefulness of the method.

Optical spectroscopic techniques are label-free, non-destructive and more accurate in early and rapid detection of diseases and enable us to use them as a large-scale screening tool for citrus diseases. Over the past decade, there has been an increase in academic research focused on utilizing spectroscopic techniques for identifying diseases in citrus plants. Plant disease identification is an important part of crop plant management in agriculture and horticulture. In the case of citrus disease infections, the pathogen can remain in a latent form inside the plant. However, the timing and extent of its activity can vary, making it challenging to predict when symptoms will become visible. This variability not only makes managing citrus diseases, including HLB, difficult but also offers an excellent opportunity for the development and advancement of optical spectroscopy-based biophotonics tools and sensors. Another challenge that still needs to be addressed is the scenario in which infestations of more than one disease, also known as multiple disease complexes or co-infections, exist in a plant, presenting a significant challenge in detection. The capability of early detection of stress and infections using the light-based spectroscopic tools is advantageous for farmers, producers and horticulturist as it enables prompt intervention to prevent and early control the spread of infection and consequently minimize the loss and degradation in citrus quality. Thus, the information obtained from spectroscopy can be used to make management decisions, including the implementation of suitable control measures to minimize the impact of various diseases on citrus plants.

Advances in light sources, detector technologies, chemometrics and image processing methods will improve the ability of optics-based methods to diagnose citrus diseases and will support to ensure a better future for the citrus farming and precision agriculture. An integration of spectroscopic techniques with artificial learning algorithms opens a new field of study to rapidly detect diseases in citrus plants on an agro-industrial and commercial scale. These rapid, label-free and portable spectroscopic techniques could potentially be utilized on a large scale in agriculture technology to detect citrus disease at an initial growth stage. Furthermore, the integration of optical spectroscopic techniques with omics technologies, such as transcriptomics, metabolomics, and genomics in association with a big data analysis approach, can provide a holistic view of citrus disease initiation and progression in plants. This approach can elucidate complex plant response to environmental changes and stressors in a better way. Moreover, when employing a multimodal approach for analysis, it is imperative to explore the alterations in substances during the development of citrus diseases and determine which substances play the most significant role in diagnosing citrus diseases.

**Funding:** This research was supported by the Science and Engineering Research Board (grant number SRG/2020/00220) and, Ramalingaswami Re-entry Fellowship, Department of Biotechnology, Government of India

**Conflicts of interest/Competing interests:** The authors declare no conflict of interest.

**Data availability:** None

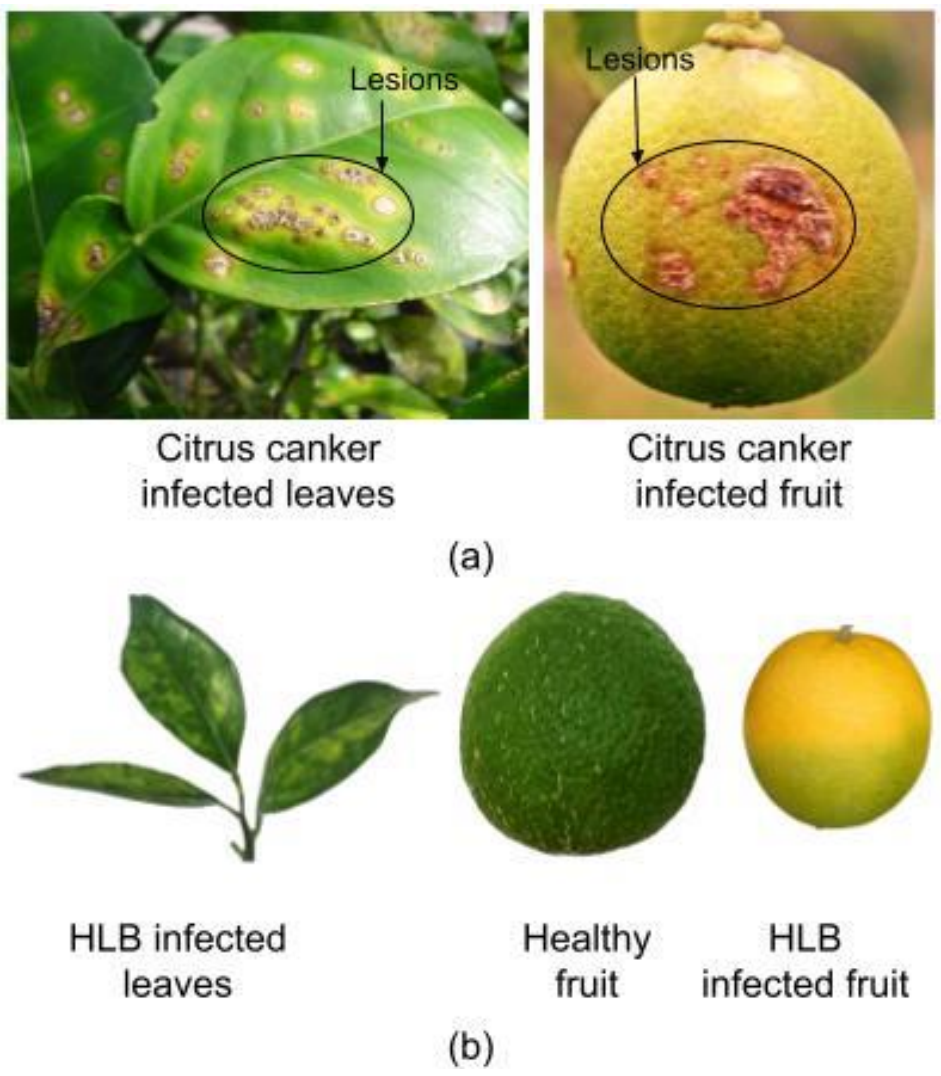

Fig. 1 Symptoms of citrus canker and HLB infection in leaves and fruits. (a) The lesions associated with citrus canker exhibit a round or oval shape, featuring a distinctive raised center surrounded by a water-soaked margin. (b) The HLB infection demonstrates the symptoms such as yellowing of leaf veins with blotchy mottling, as well as lopsided, underdeveloped and premature drop of the fruit.

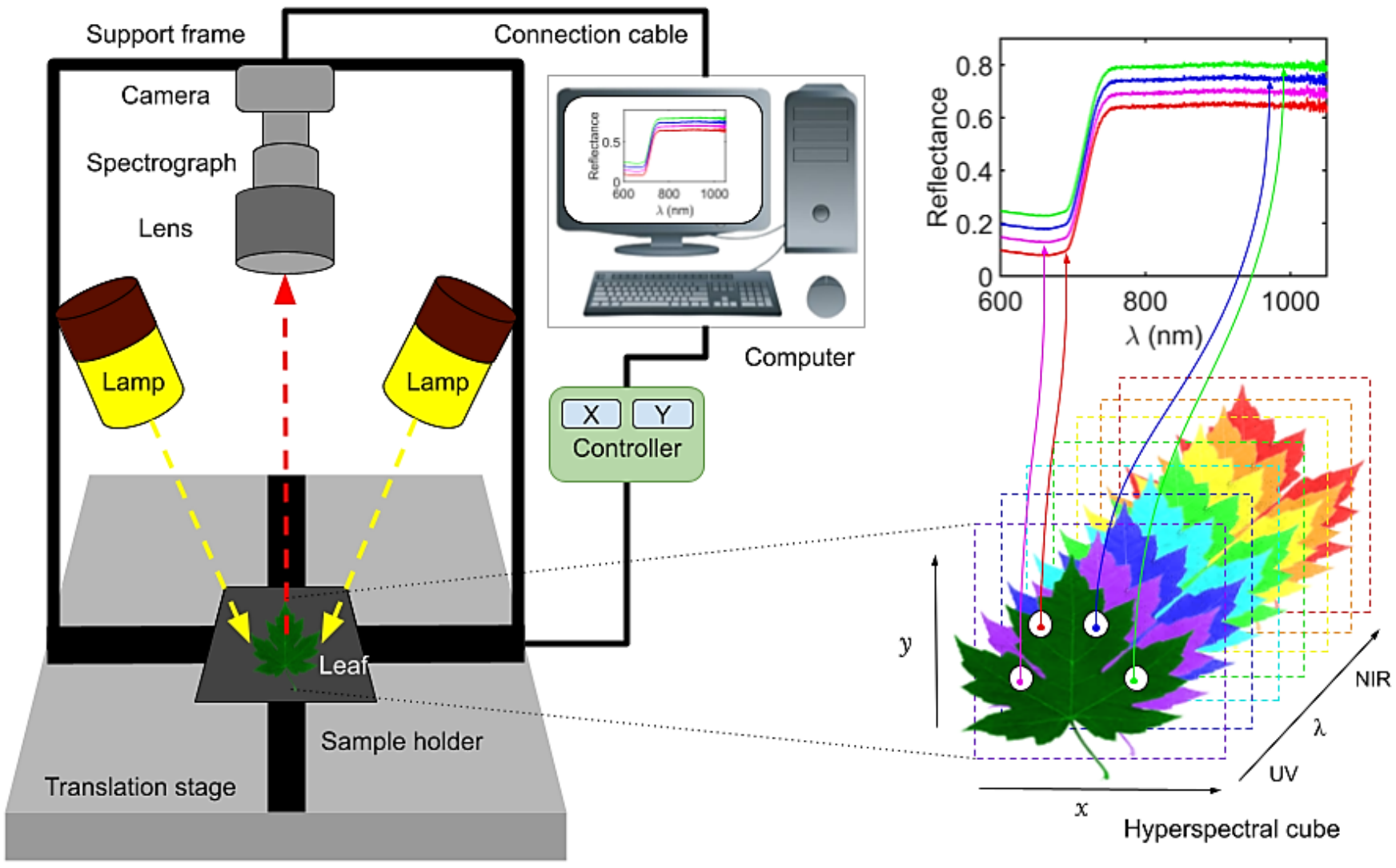

Fig. 2 A schematic diagram of hyperspectral camera and indoor set-up to capture hyperspectral image of a leaf.

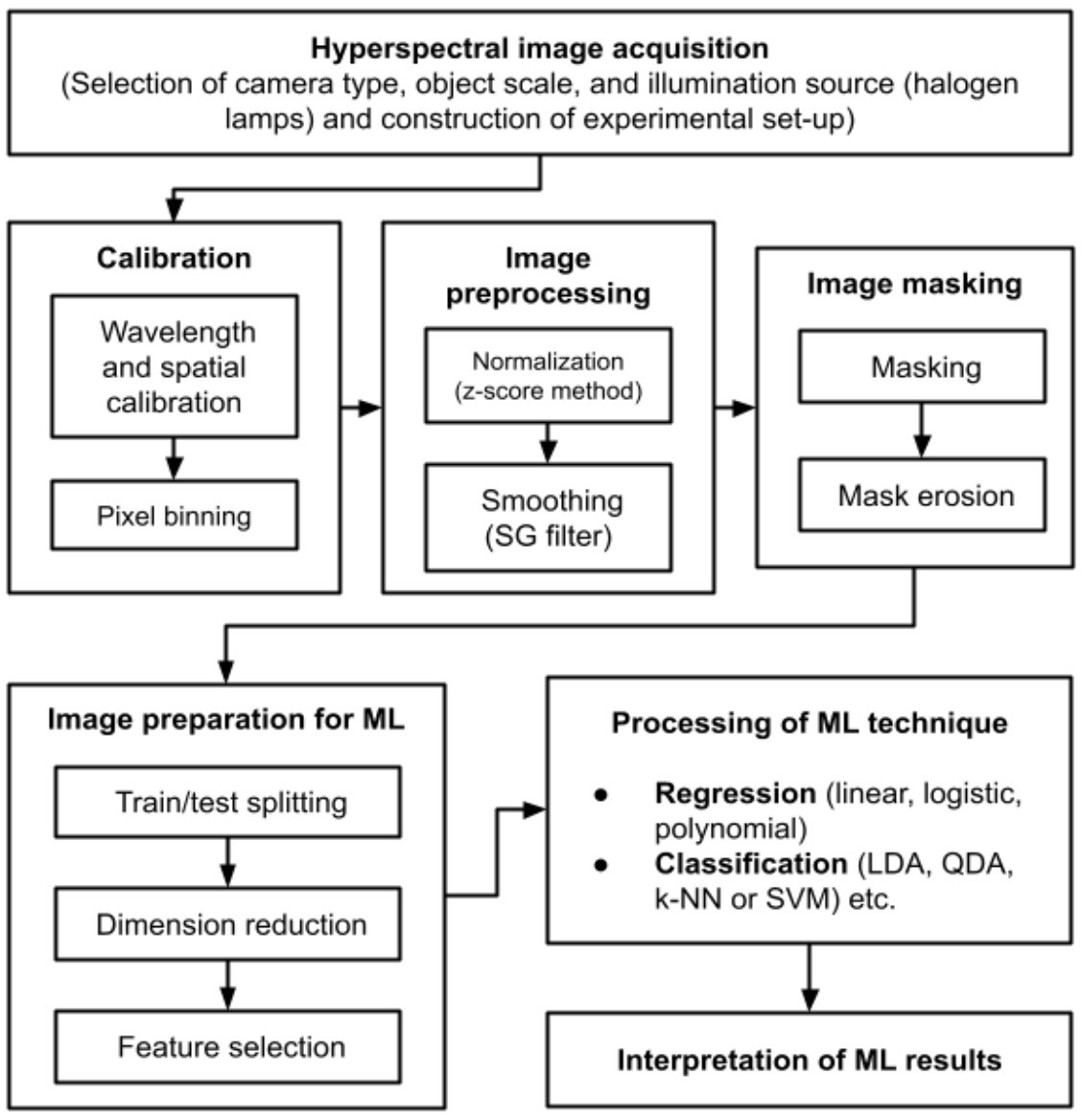

Fig. 3 The generalized hyperspectral workflow encompasses various steps, starting from the acquisition of hyperspectral images to the identification of regions of interest in the captured images.

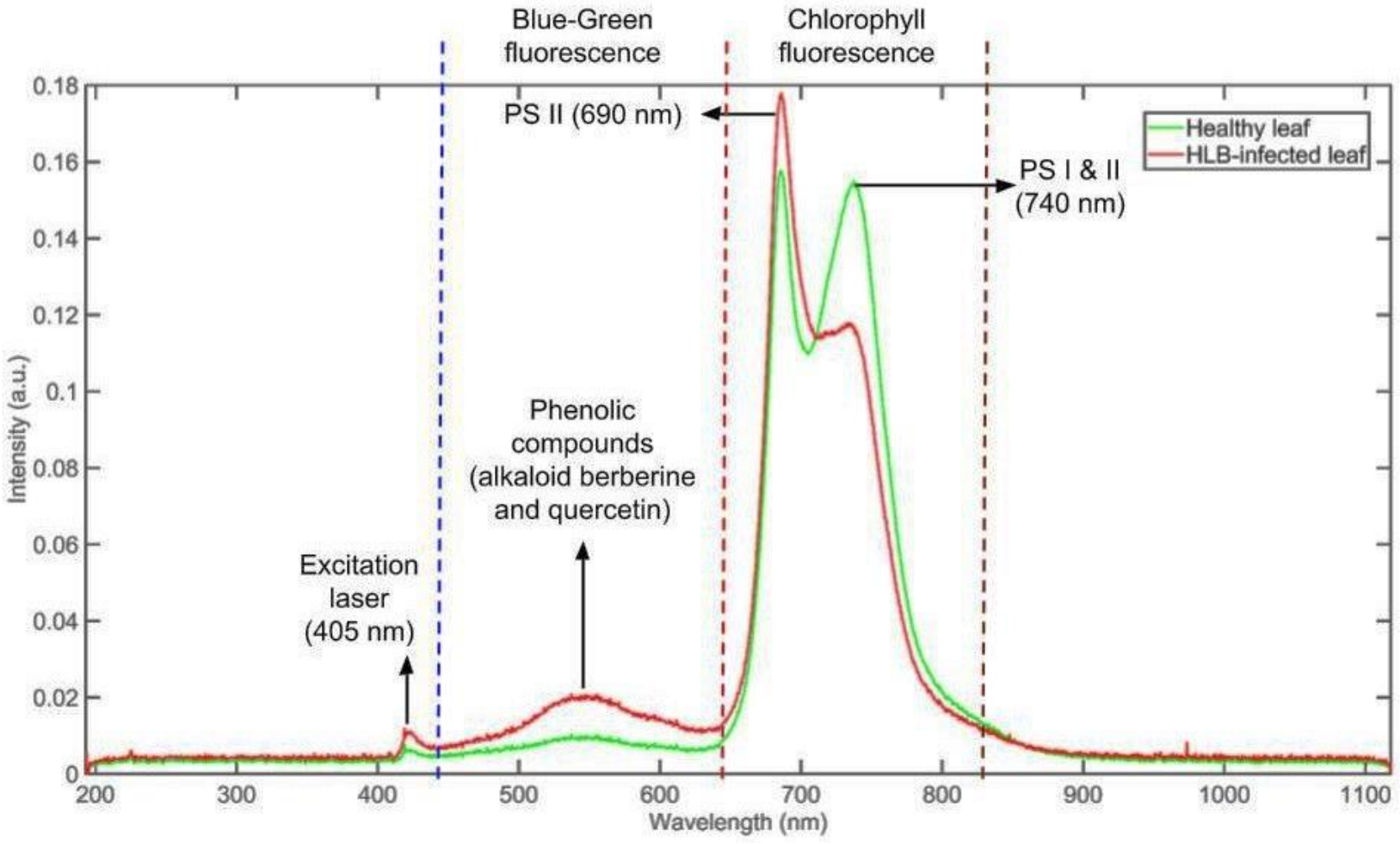

Fig. 4 The fluorescence spectra were measured from the adaxial surface of healthy and HLB-infected citrus leaves using a 405 nm excitation laser.

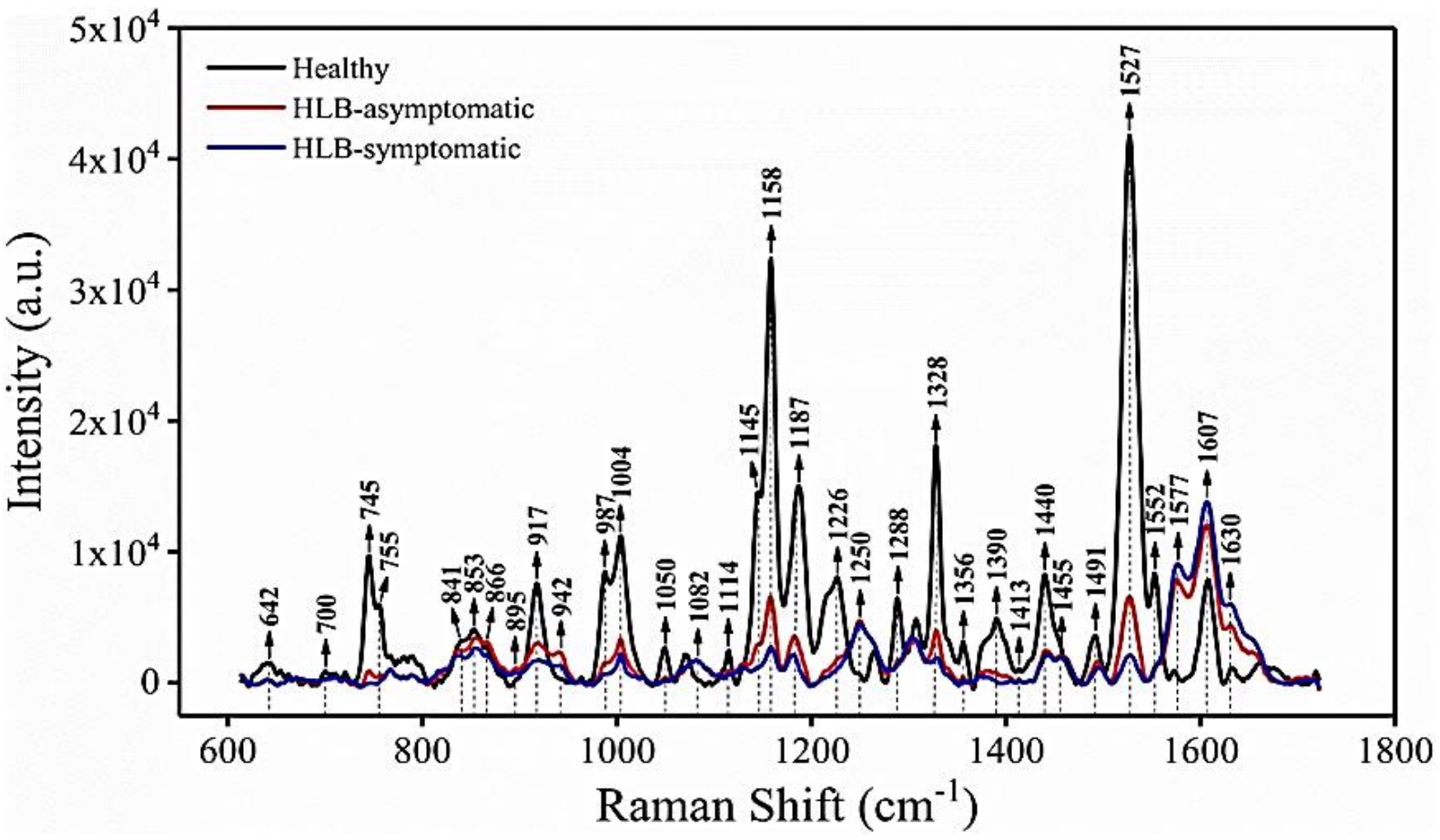

Fig. 5 Raman spectra of healthy and HLB-infected (asymptomatic and symptomatic) citrus leaves [114] [reprinted and adapted with permission].

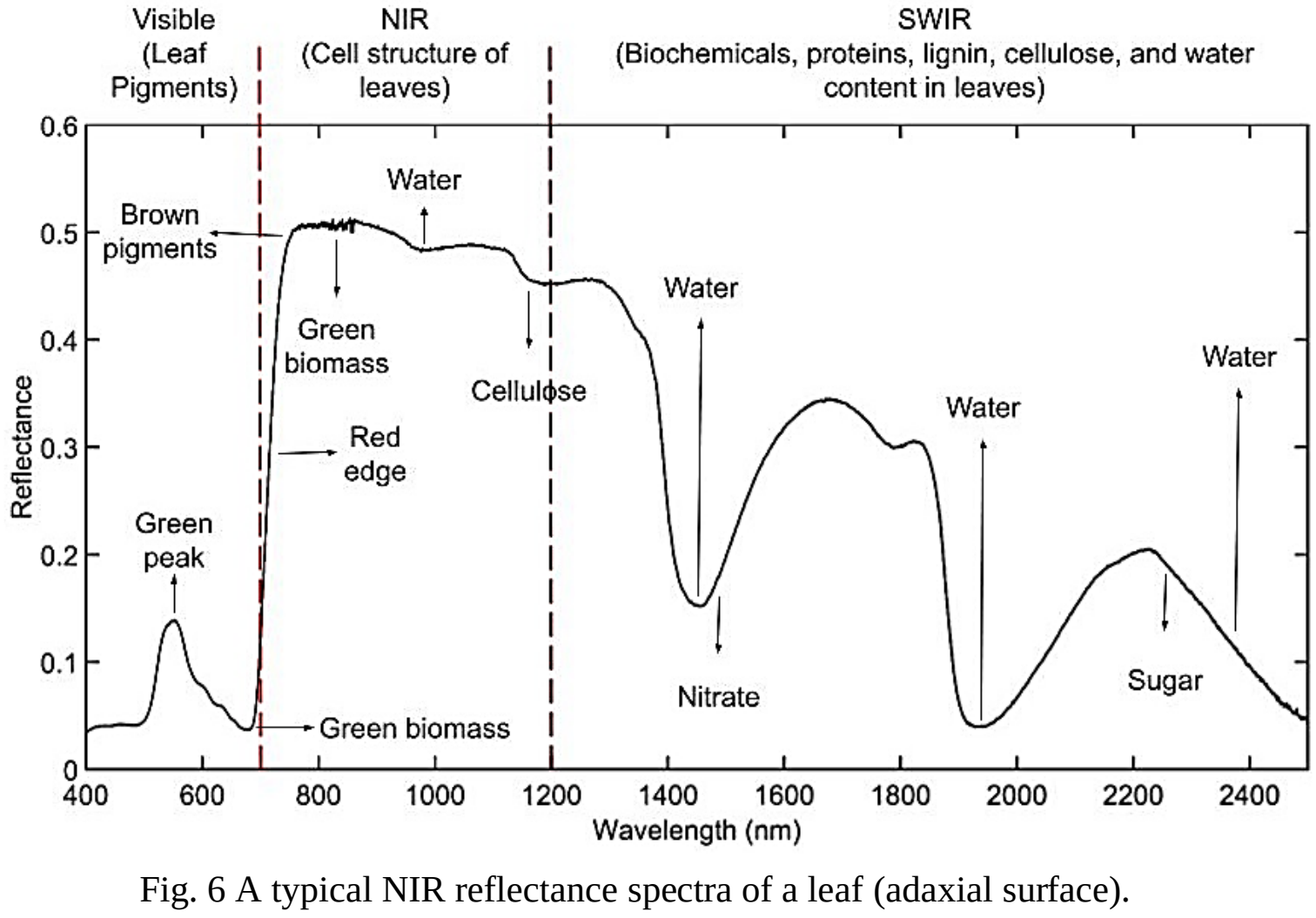

Fig. 6 A typical NIR reflectance spectra of a leaf (adaxial surface).

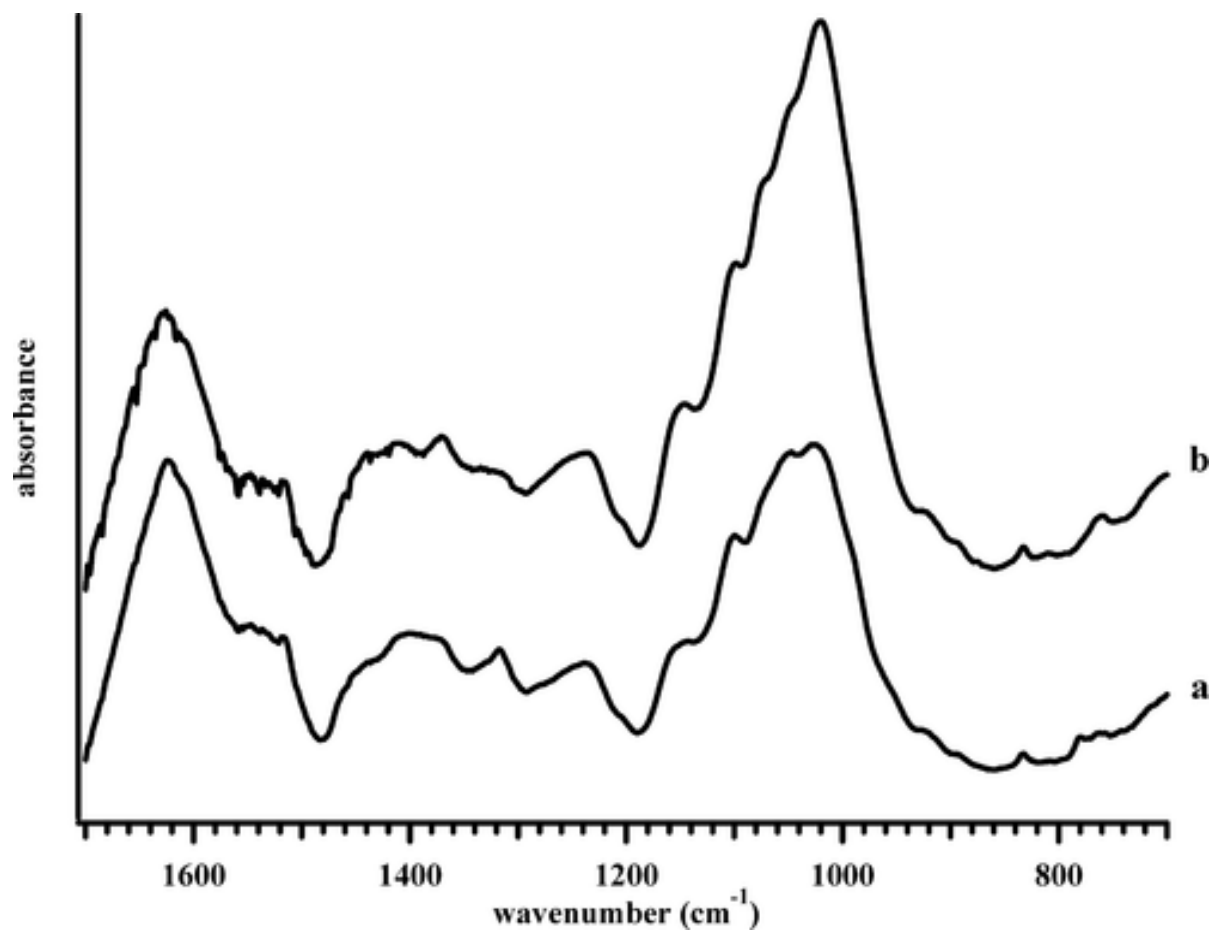

Fig. 7 The FTIR spectra of (a) HLB negative and (b) HLB positive citrus leaves [28] [reprinted and adapted with permission].

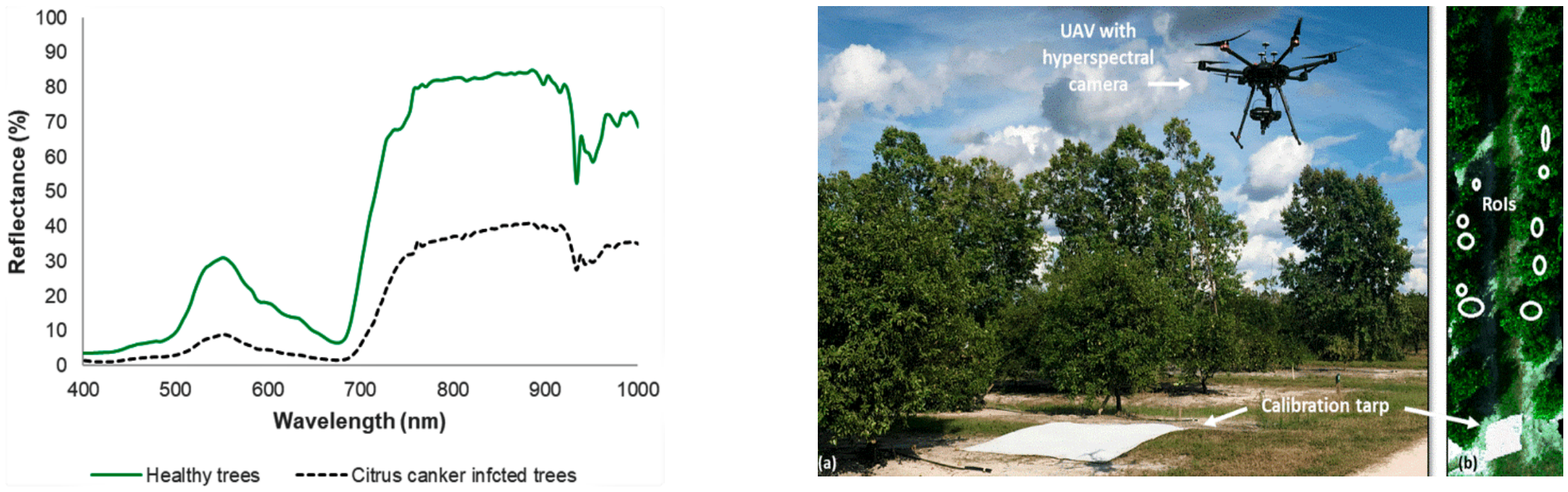

(a) (b)

Fig. 8 Reflectance spectra of healthy and citrus canker infected plants acquired from UAV-based hyperspectral imaging system. The hyperspectral data collection system utilizing UAV technology captures images of citrus canker-infected plants within the orchard. (a) UAV hyperspectral imaging system and (b) regions of interest were chosen from the plants affected by citrus canker [11].